\documentclass{aastex}
\begin{document}

\title{A microscopic derivation of Special Relativity: \\
simple harmonic oscillations of a moving space-time lattice}

\author{Richard~Lieu$\,^{1}$}
\affil{\(^{\scriptstyle 1} \){Department of Physics, University of Alabama,
Huntsville, AL 35899, U.S.A.}
}

\begin{abstract}
The starting point of the theory of Special Relativity$^1$ is the Lorentz
transformation, which in essence describes the lack of absolute
measurements of space and time.  These effects came about when one applies
the Second Relativity Postulate (which states
that the speed of light is a universal
constant) to inertial observers.  
Here I demonstrate that there is a very elegant way of
explaining how exactly
nature enforces Special Relativity, which compels us to conclude that
Einstein's theory necessitated quantization of space and time.
The model proposes that microscopically the structure of space-time
is analogous to a crystal which
consists of lattice points or `tickmarks' (for measurements)
connected by identical `elastic springs'.
When at rest the `springs' are at their
natural states.  When set in
motion, however, the lattice
vibrates in a manner described 
by Einstein's theory of
the heat capacity of solids, with consequent widening of the
`tickmarks' because the root-mean-square separation now increases.
I associate
a vibration temperature $T$ with the speed of motion $v$ via the
fundamental postulate of this theory, viz. the relation
$\frac{v^2}{c^2} = e^{-\frac{\epsilon}{kT}}$ where 
$\epsilon$ is a quantum of energy
of the lattice harmonic oscillator.
A moving observer who measures distances and time intervals with
such a vibrating lattice obtains
results which are precisely those given by
the Lorentz transformation.  Apart from its
obvious beauty, this approach provides many new 
prospects in understanding space and time.  For example, 
an important consequence of the model
is the equation $\epsilon = \kappa x_o^2$, where $x_o$ is the
basic `quantum of space' and $\kappa$ is the spring constant which
holds together the lattice.  Thus space-time, like mass, has
an equivalence with energy.
\end{abstract}
\hspace{1cm} \\

In Special Relativity theory the distance between two points
is not absolute.  If an inertial observer $\Sigma'$ measures
{\it simultaneously} the positions of the ends of a uniformly moving rod
to obtain the rod length, and if the rod is oriented
parallel to its velocity ${\bf v}$, the result will be smaller than that
of another inertial observer $\Sigma$ who measures the length from a frame of
reference with respect to which the rod is at rest.
The difference between the two data values is
the Lorentz factor $\gamma = (1 - v^2/c^2)^{-\frac{1}{2}}$,
which means the length obtained by $\Sigma'$ is much smaller
in the limit $v \rightarrow c$.  The same conclusion applies
to the measurement of time.  Thus if $\Sigma'$ witnesses a time
difference between the decay of two moving elementary particles,
both having the same velocity
${\bf v}$ and disintegrated in the {\it same} spatial position,
the result is smaller than that of $\Sigma$, who measures decay times
from a reference frame which does not move with respect to either
particles.

The above manner of formulating the two paramount results of Special
Relativity considers (a) the effects of motion on position measurements
at the same time, and (b) time measurements at the same position.  Once
each phenomena is understood without interference
from the other, changes in space and
time can be superposed to form a general Lorentz transformation.
Einstein originally arrived at this transformation by invoking the Second
Relativity Postulate.
I propose in this {\it Letter} a new more fruitful approach which
demonstrates how Lorentz transformation is realized by a dynamic
space-time
lattice.  We begin with
a treatment of the relativity of length, i.e. situation (a).
The treatment of time then
follows immediately by considering the time lattice as being the same
as that of space, except re-scaled by the speed of light $c$ to form
a different dimension. 

The measurement ruler of $\Sigma$ is calibrated according to the
topmost part of Figure 1, and is used to determine the length of the rod
drawn immediately beneath.  A minimum unit of length is assumed to
exist (i.e. space is quantized), which we designate as $x_o$, and is
also the distance between any pair of black dots 
(hereafter referred to as lattice points) on the ruler.  The
length of the rod is then expressed in units of $x_o$ (such a regular
lattice is not a necessary part of the theory, but for simplicity of
argument we shall assume it).
If the ruler of a moving observer measures a smaller length for the same
rod, this is equivalent to having a larger quantum of distance
for $\Sigma'$, i.e. the
lattice points of $\Sigma'$ are more widely spaced, as shown in the
lower half of Figure 1.  Specifically the length contraction effect
will be achieved if, for $\Sigma'$:
\begin{equation}
x_o \rightarrow x_o \left(\frac{1}{1 - \frac{v^2}{c^2}} \right)^{\frac{1}{2}}
\end{equation}
where ${\bf v}$, the velocity of $\Sigma'$ relative to $\Sigma$, is
parallel to the length of the rod.

For the lattice of $\Sigma'$ every lattice point
moves with velocity ${\bf v}$ like all others, so why would this lead to
a widening of the lattice ?  The question arises only because
we consider motion of the lattice as a rigid body.  It is entirely
possible that microscopically the space lattice points are connected
by springs, and a moving lattice necessarily vibrates:
the amplitude of oscillation increasing
with $v$.  Henceforth I shall address the previous sentence 
(to be made even more precise in the enusing discussions) as
the {\it Postulate of Lattice Space-Time}.
Now there is an effective increase
in the lattice spacing $x$ because,
while the mean separation $<x>$ remains at $x_o$, the 
{\it root mean square} $x_{rms} = \sqrt{<x^2>}$ is larger.
Indeed, an
interesting aspect of Equ (1) is that the right side has
the form $x_o \sqrt{N}$ where $N \geq 1$, which reminds one of random
walk.  Moreover, $N$ has a denominator reminiscent of quantum statistical
mechanics.

To explain the Lorentz transformation, I would then begin
with a harmonic oscillator.  We write $x = x_o + x_1$, with
$x_1$ representing pure sinusoidal motion, implying $<x_1> = 0$,
and the lattice spacing  may be written as:
\begin{equation}
x_{rms} = x_o \left(1 + \frac{<x_1^2>}{x_o^2} \right)^{\frac{1}{2}}
\end{equation}
Next, we assume a classical
oscillator and, to quantify the forementioned Postulate
of Lattice Space-Time, we make the reasonable conjecture that
$\sqrt{<x_1^2>} \propto v = \alpha v$ (meaning of course that
$\sqrt{<\dot{x}_1^2>}$ is also $\propto v$; such a
relation applies, e.g. to a uniformly moving lattice which started from
rest as a result of an external impulse).  
Equ (2) now reads:
\begin{equation}
x_{rms} = x_o \left(1 + \frac{\alpha^2 v^2}{x_o^2} \right)^{\frac{1}{2}}
\end{equation}
We can therefore derive the length contraction of Equ (1) 
in the non-relativistic limit of $v \ll c$ if in Equ (3) we
set $x_o^2/\alpha^2 = c^2$.  However, 
apart from the obvious lack of elegance and
completeness, the model is at best heuristic.
An exact treatment requires quantum mechanics, when the energy of
a harmonic oscillator becomes $E = (n + 1/2) \epsilon$ where $\epsilon$
is the basic quantum of the system.   Once the complete theory is
in place, we will see that the case of $v \ll c$ is not even classical.

The proposed model works successfully if it allows the
lattice oscillators to have varying energies $E$, but the 
mean energy and degree of
variation is governed 
by a temperature $T$ which replaces $E$ as the monitoring
parameter of  the overall vibration level (i.e. $T$
increases with 
$v$).  The model used for such a
thermal quantum lattice is Einstein's theory for the heat
capacity of solids.  Here the reader is
reminded of the 
Partition function and mean energy per oscillator.  Ignoring the
zero point energy they are, 
respectively:
\begin{equation}
Z = \frac{1}{1 - e^{-\frac{\epsilon}{kT}}}
\end{equation}
and $\bar{E} = \epsilon e^{-\frac{\epsilon}{kT}} \times Z$.
Thus if we define $\kappa$ as the lattice spring constant, 
and refer the average $<x_1^2>$ to its quantum expectation value, we may
write 
\begin{equation}
<x_1^2> =  \bar{E}/\kappa =
\frac{\epsilon Z}{\kappa} e^{-\frac{\epsilon}{kT}}
\end{equation}
When Equ (4) and (5) are substituted into
Equ (2), the result is in close resemblance
with that of length contraction (Equ (1)).  

In fact, the analogy is
striking, and provides clear guidelines on how precisely $T$ is
related to $v$.  Thus, to quantitatively complete the Postulate of
Lattice Space-Time, I propose this relation as: 
\begin{equation}
\frac{v^2}{c^2} = e^{-\frac{\epsilon}{kT}}
\end{equation}
Use of Equ (4), (5), and (6) gives:
\begin{equation}
<x_1^2> = \frac{\epsilon}{\kappa} \frac{\frac{v^2}{c^2}}{1 - \frac{v^2}{c^2}}
\end{equation}
Equ (2) and (7) now combine to read:
\begin{equation}
x_{rms} = x_o \left(1 + 
\frac{\epsilon}{\kappa x_o^2} \frac{\frac{v^2}{c^2}}{1 - \frac{v^2}{c^2}}
\right)^{\frac{1}{2}}
\end{equation}
The coincidence in form between Equ (8) and Equ (1), with the former
being based on a totally natural model which involves no manipulation
of terms, renders it extremely difficult to draw conclusions
other than the one which says that we are indeed confronted with a
microscopic realization of Relativity.  In fact,
Equ (8) agrees perfectly with Equ (1) when the following simple relationship
between the three fundamental parameters of the space-time lattice holds:
\begin{equation}
\epsilon = \kappa x_o^2
\end{equation}
A similar equation which gives no extra information applies to the
time lattice, and is responsible for the relativity of time described 
earlier.
This is because in Lorentz transformation time behaves exactly like
space, with the substitution $x = ct$.  Thus time and space are
controlled by the same underlying lattice.
 
In conclusion , Special Relativity already revealed that even the
arrangement of the smallest quanta of our space-time fabric
is highly ordered.  However, one must
also consider Equ (9), which
implies that the space-time `crystal' has an energy equivalence.
A more immediate consequence is the quantization of 
the Lorentz factor $\gamma$
at low $v$.  Could this be responsible for the discreteness in the mass
of elementary particles ?
The reported findings
may also be of relevance to understanding
astrophysical processes, including (and
especially) those of the early universe.  Finally, it opens the
question as to how the notion of space-time microstates presented here
could facilitate further understanding and development of 
the theory of General
Relativity.

I am indebted to Dr Massimilano Bonamente for suggesting a
harmonic oscillator model for the space-time lattice of a
moving observer.

\section*{Reference}
 
[1] Einstein, A., 1905, {\it Annalen der Physik}, {\bf 18}, 891.

\newpage
 
\section*{Figure Caption}
 
The space-time lattice of stationary ($\Sigma$)
and moving ($\Sigma'$) observers are
illustrated here for the case of distance measurements.  The
`tickmarks' of the ruler of $\Sigma$ are marked as the topmost
set of black dots.  The rod to be measured is the short bar
immediately beneath, and is at rest with respect to $\Sigma$.
Observer $\Sigma'$ measures the length of this rod while in
motion, by simultaneously acquiring data on the positions of
the front and rear end of the rod.  It is postulated that
effectively $\Sigma'$ is using a moving set of `tickmarks',
and if microscopically these are connected by `elastic springs'
which can extend while in motion, the `tickmarks' widen as
depicted in the lower half of the diagram.  Consequently
$\Sigma'$ obtains a smaller value for the length of the rod.

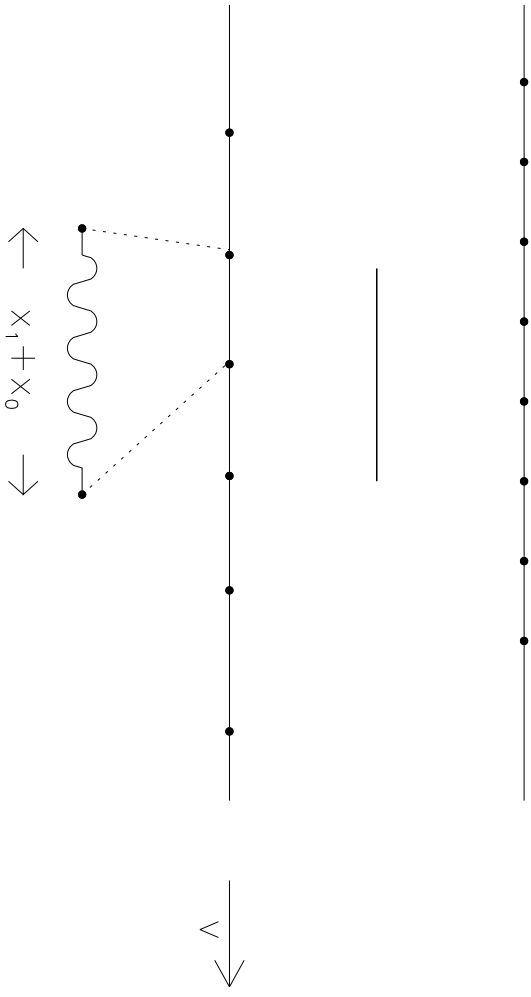
\end{document}